\documentstyle[aps]{revtex}
\begin{document}


\title{\begin{bf}
Problems with Extraction of the Nucleon to Delta(1232)
Photonic Amplitudes
\end{bf}}

\author{R.~M.~Davidson$^a$, Nimai C.~Mukhopadhyay$^a$ and
M.~S.~Pierce$^{a}$\\
R.~A.~Arndt$^b$, I.~I.~Strakovsky$^{b,c}$ and R.~L.~Workman$^b$}
 
\address{
$^a$Department of Physics, Applied Physics and Astronomy\\
Rensselaer Polytechnic Institute, Troy, NY 12180-3590\\
$^b$Department of Physics\\
Virginia Polytechnic Institute and State University\\
Blacksburg, VA 24061\\
$^c$Department of Physics\\
The George Washington University\\
Washington, DC 20052}

\maketitle

\begin{abstract}
We investigate the model dependence and the importance of choice of database
in extracting the {\it physical}
nucleon-Delta(1232) electromagnetic transition amplitudes, of interest to
QCD and baryon structure, from the
pion photoproduction observables. 
The model dependence is found
to be much smaller than the range of values obtained when different
datasets are fitted. In addition, some
inconsistencies in the current database are discovered, and
their affect on the extracted transition amplitudes is discussed.
\end{abstract}


The importance of electromagnetic amplitudes in the
nucleon to Delta(1232) transition as a test of our understanding
of the underlying hadron dynamics
has been known for decades \cite{ncm}.
In the SU(6) symmetry limit of the quark model \cite{bec},
the only surviving amplitude is the transverse magnetic dipole ($M1$).
The other allowed electromagnetic transitions, the electric and scalar
(longitudinal) quadrupole, $E2$ and $S2$ (L2) amplitudes respectively,
vanish identically in this symmetry limit. In more realistic quark models
\cite{quark}, the $M1$ amplitude for real photon transitions is predicted
to be about 30\% lower than what is extracted \cite{da86,rpi,vpi}
from the data, while the $E2$ amplitude is considerably
smaller in magnitude than the extracted value. Other models of hadrons,
such as bags \cite{bag}, Skyrmions \cite{skyrm}, other solitons \cite{sol}
and the more rigorous QCD approach on the lattice \cite{lat}, all give
varying values for these amplitudes, none in perfect agreement with the
phenomenologically inferred values \cite{da86,vpi}. A relevant quantity
that reflects this theoretical interest is the ratio of $E2$ to $M1$ at the
K-matrix pole (where the $\pi N$ phase shift in the J=T=3/2 (or 3,3)
channel goes through 90$^0$). At the real photon point, $Q^2$ = 0, this
ratio is small and negative, while at large $Q^2$, in the domain of
perturbative QCD, this ratio is expected to be positive and unity \cite{carl}.

Current research on this topic has theorists struggling
to understand why theoretical predictions at the real photon point
fall short of the values inferred from experiment. For example, introducing
explicit pion degrees of freedom may be the solution to this long-standing
problem. An early work \cite{kon}
based on a simple pion emission model found that
the $M1$ amplitude is largely unaffected by the pion cloud. More
recent works on this topic are divided. The work of Sato and Lee \cite{sato},
who
use an effective Hamiltonian method, find that the pion cloud is important
for the $M1$ amplitude. On the other hand, a different work \cite{alf}
in which the one-pion exchange potential is included in the quark model
finds that the $M1$ amplitude is largely unaffected by the pionic degrees
of freedom. This latter work also points out the importance of two-body
currents in the $E2$ amplitude.

At the same time,
a vigorous program has been launched by both experimentalists
and phenomenologists to determine them more accurately by doing better
experiments and analyses.
Very recently, the Mainz \cite{beck} and the
BNL LEGS \cite{bnl} groups have released newly measured
differential cross section (d$\sigma$/d$\Omega$) and   
photon asymmetry ($\Sigma$) data in the Delta(1232) resonance region.
These new data have caused tremendous interest among the phenomenologists,
and the theorists from RPI \cite{rpi}, VPI \cite{vpi} 
and Mainz \cite{mainz} have produced their own analyses of these data.
These analyses have found the
$E2/M1$ ratio ranging from $-(1.5\pm 0.5)\%$ \cite{vpi} to
$-(3.2\pm 0.2)\%$ \cite{rpi}, a range of -1\% to -3.4\%. Given that
$E2$ is so small compared to $M1$, it is quite satisfactory to have
its value narrowed down to this range. However, it is not satisfactory
that different analyses do not agree, within their quoted error bars,
on the value of $E2/M1$. One reasonable explanation of this discrepancy
is that the various authors have underestimated their errors. Are there
any other explanations for the disagreement amongst the various
analyses?

This brings us to the central issue of this work: finding the origin of
this numerical spread of the $E2/M1$ ratio and what can be done to
substantially reduce this spread.
One might suspect \cite{wil}
that this spread is coming from the theoretical model 
dependence of various analyses. However, we will show that the model dependence
in extracting the {\it physical} photo-decay amplitudes is \begin{it}actually
much smaller than the effect of different choices for the
databases that various authors have made\end{it} \cite{rpi,vpi,mainz}.
Indeed, if a complete set of experiments were available at the K-matrix
pole position, then the multipoles could be determined model independently,
and thus there would be no model dependence in determining the K-matrix residues
\cite{dav}. As a complete set of experiments does not yet exist,
model
dependence arises in the multipoles due to what one assumes
about the structure of the photoproduction amplitude.

We construct test datasets of pertinent observables in the energy region
between the pion photoproduction threshold and the onset of the two pion
threshold, beyond which implementation of unitarity is difficult. We then
apply two distinct theoretical approaches of extracting the
resonant amplitudes to the same dataset, to
gauge the degree of model dependence in the extraction
of the resonant amplitudes.
Both of these approaches
provide a generally good description of the photoproduction data, but
differ significantly in their physics input.
One of these two is the effective 
Lagrangian approach \cite{da86,rpi,dmw} which satisfies the constraints of
chiral symmetry, unitarity and crossing symmetry.
The other is
a more phenomenological energy-dependent (global) multipole analysis
\cite{vpi}. 

The first approach, hereafter called the RPI analysis \cite{da86,rpi,dmw},
is based on an effective Lagrangian \cite{dmw}
containing the pseudovector (PV) nucleon Born terms, s and u channel
$\Delta$ exchange and t-channel $\rho$ and $\omega$ exchanges. It
contains five parameters that are determined by fits to the data: two
gauge couplings describing the $M1$ and $E2$ amplitudes and
three off-shell parameters
related to ambiguities in the relativistic spin-3/2 Lagrangian \cite{ben}.
In this
approach, the d and higher-$l$ multipoles are dominated by the nucleon Born
terms, i.e., effectively only the s and p wave multipoles are allowed to vary.
In addition, not all the s and p wave multipoles are independent as the
model contains only five parameters. While the resonance
contribution is not {\it a priori} constrained, the background is.
In fact,
for the most part the background is dominated by the PV nucleon Born terms.

The second approach, which we call the VPI analysis \cite{vpi}, has
the energy dependence of the s, p and d-waves
multipoles parameterized and unitarity is implemented via a K-matrix
approach. The higher-$l$ multipoles are assumed to be given by the
nucleon Born terms, and the parameters, including $M1$ and $E2$,
are determined by a fit to the data. Thus, apart from the $l$ $>$ 2
multipoles, the background and resonance contributions are not {\it a priori}
constrained in this approach. We shall repeat our analysis for two
different test databases to demonstrate the role of the choice of the database
in extracting resonance parameters.

For this analysis, we need a standard set of definitions for resonance
parameters \cite{dav,dmw}.
For the K-matrix definition of the resonance photo-couplings, we have \cite{dav}
\begin{eqnarray}
M1 = {\rm Im} M_{1+}^3 A \; , \nonumber \\
E2 = {\rm Im} E_{1+}^3 A \; ,
\end{eqnarray}
where
\begin{equation}\label{adef}
A = \sqrt{ {8\pi Wq\Gamma_{\Delta} \over 3MK }} \; ,
\end{equation}
and all kinematical quantities are evaluated at the cm energy where the
phase shift in the isospin 3/2, spin 3/2 (3,3) channel is 90$^0$.
In (\ref{adef}), $M$ is the nucleon mass,
$\Gamma_{\Delta}$ is the Delta width at the K-matrix pole \cite{dav},
$q$ is the
pion three-momentum , and $K$ is the photon three-momentum.

At the T-matrix pole, we adopt a definition of the photo-couplings which
is similar to that used at the K-matrix pole. First, consider pion-nucleon
elastic scattering in the 3,3 channel. Below energies where inelasticities
become important, the T-matrix is of the well-known elastic form
\begin{equation}\label{tmat}
T= \sin \delta {\rm e}^{i\delta} \; .
\end{equation}
If we now define the energy dependence of the Delta width, $\Gamma (W)$,
to be given by
\begin{equation}
\tan \delta = { \Gamma (W) /2 \over W_R - W} \; ,
\end{equation}
then  the T-matrix in the 3,3 channel becomes, without loss of generality,
\begin{equation}\label{tmat3}
T = { \Gamma (W)/2 \over W_R -W -i \Gamma (W) /2 } \; .
\end{equation}
Note that in (\ref{tmat3}),
$W_R$ is the energy where the 3,3 phase shift is 90$^0$. In the
literature, one often finds the T-matrix in the 3,3 channel written
in the above form (\ref{tmat3}) {\it plus} an additional background
contribution. However, the background contribution may be absorbed
into the definition of $\Gamma (W)$.

In the vicinity of the T-matrix pole, $W_T$, we have
\begin{equation}
\Gamma (W) \approx \Gamma (W_T) + (W-W_T) \Gamma ' (W_T) + \cdots \; ,
\end{equation}
and by using the standard definition of $W_T$,
\begin{equation}
W_T = W_R - i \Gamma (W_T)/2 \; ,
\end{equation}
we obtain, in the vicinity of $W_T$,
\begin{equation}
T \approx { \Gamma (W_T)/2 \over (W_T -W) [ 1+i \Gamma ' (W_T)/2 ] } \; .
\end{equation}
Therefore, the residue at the T-pole is
\begin{equation}
{\rm Res} T \equiv R_S =
{ \Gamma (W_T)/2 \over  1+i \Gamma ' (W_T)/2  } \; .
\end{equation}
{}From (7), one trivially obtains
\begin{equation}\label{gam}
\Gamma (W_T) = 2i(W_T -W_R) \; ,
\end{equation}
and we also define the T-matrix
width by $\Gamma_T$ = Re$[(\Gamma (W_T))]$.

Turning to photoproduction, the $M_{1+}$ multipole in the T=3/2
channel, $M_{1+}^3$, may be written as
\begin{equation}\label{m1res}
M_{1+}^3 = M1(W) \sqrt{ { 3KM \over 8\pi W q \Gamma (W) }}
{ \Gamma (W)/2 \over W_R -W -i \Gamma (W) /2 } \; .
\end{equation}
In terms of the residue of $M_{1+}^3$ at the T-matrix pole, we thus
obtain
\begin{equation}\label{m1}
M1 (W_T) = { {\rm Res} M_{1+}^3 \over R_S } \sqrt{ { 8\pi W_T q
\Gamma (W_T) \over 3KM}} \; ,
\end{equation}
where $\Gamma (W_T)$ is given by (\ref{gam}). The relevant expression for
$E2 (W_T)$ can be found from (\ref{m1}) by replacing $M1$ with $E2$
and $M_{1+}^3$ with $E_{1+}^3$.
To make connection with the quantity $re^{i\phi}$ introduced by the
Mainz group \cite{mainz}, we obtain
\begin{equation}
{\rm Res}(M_{1+}^3) = re^{i\phi}\Gamma_T \; .
\end{equation}

In the RPI approach, Res$(M_{1+}^3)$ may be exactly calculated from
Eq.~(55) of Ref.~\cite{dmw} once the parameters have been fitted to the
data. However, in other approaches it is necessary to fit some functional
form to the multipoles and extrapolate to the T-matrix pole. Although
countless extrapolation functions can be used, we expect, based on results
from $\pi N$ scattering \cite{licht}, that the extrapolation is not very
sensitive to the particular parameterization of the multipoles. Thus,
we adopt the simple form
\begin{equation}\label{extrap}
f(W) = { \mu^2 \over qK }\sin \delta e^{i\delta} (A+BX+CX^2 ) \; ,
\end{equation}
where $X$ = $(W-M_R )/M_R$, $\mu$ is the pion mass,
and the constants $A,B,C$ are determined
from a fit to the multipole. This form, (\ref{extrap}), provides an
accurate description of both the $M_{1+}^3$ and $E_{1+}^3$ multipoles
in the $W$ range from 1180 to 1250 MeV. The T-matrix residues are
given by
\begin{equation}
{\rm Res}(M_{1+}^3) = R_S { \mu^2 \over qK} (A+BX+CX^2) \; ,
\end{equation}
where all kinematical quantities are evaluated at $W_T$. A similar
equation holds for Res$(E_{1+}^3)$. For the RPI approach, the results
obtained from the extrapolation function (\ref{extrap}) agree with the
exact results to within 1\% for the $M_{1+}^3$ and within 5\% for
the $E_{1+}^3$.

Let us now make our case that
much of the difference in the extracted $M1$ and $E2$ values, and hence the
$E2/M1$ ratios, can be traced back to
the use of different databases in the various fits. Not surprisingly, 
the recent Mainz \cite{mainz} and BNL \cite{bnl}
fits have been based largely on data 
produced at their own facilities. In order to investigate the implications
of the Mainz data on the $M1$ and $E2$ amplitudes,
the initial RPI fit \cite{rpi} 
was restricted to the Mainz data \cite{beck} over the 
resonance. The initial VPI fits \cite{vpi} were different 
in that they included the entire database. 
In order to investigate the model dependence and the dependence
on the choice of dataset
in extracting the values of the $M1$ and $E2$ amplitudes, we
have performed several fits using different, but common, datasets.
We have chosen to fit the
data in the energy region of the recent Mainz experiment, i.e., from
270 to 420 MeV. This is partially dictated by the fact that
RPI model \cite{dmw} does not
work outside the Delta region, and the VPI analysis \cite{vpi} cannot
be restricted to a narrow energy region.
Furthermore, as we are primarily interested in the
$M_{1+}^3$ and $E_{1+}^3$ multipoles, we have fitted only the proton
data, which are sufficient to isolate the isospin 3/2 multipoles.

Our first test of model dependence
is a fit which arbitrarily rejects all
pre-1980 cross section data, but keeps all
single-polarization observables. This dataset contains 836 datum points, of
which 353 correspond to $p\pi^0$ observables. In particular, there
are 140 photon asymmetry points, 91 differential cross section
points, 68 target asymmetry points, and 54 recoil polarization points.
Of the 483 $n\pi^+$ datum points, 164 are photon asymmetry points,
144 are differential cross section points, 121 are target asymmetry points,
and 54 are recoil polarization points. The effect of the few double
polarization points available in this energy region was found to be
negligible.

The results for the resonant amplitudes, $M1$ and $E2$,
in standard units of 10$^{-3}$ GeV$^{-1/2}$,
and the $E2/M1$ ratio are shown in Table 1 at the K-matrix pole
and in Table 2 at the T-matrix
pole (labelled by `Fit 1').
The results from the RPI and VPI approaches
for the dominant $M1$ amplitude are in
excellent agreement at both the K- and T-matrix
poles. Somewhat surprising is how well the RPI and VPI analyses agree on the
value of $E2$ and hence the $E2/M1$ ratio. For example, at the K-matrix
pole, the RPI analysis gives $E2/M1$ = -2.1\% while the VPI analysis gives
$E2/M1$ = -1.9\%. Based solely on the quoted systematic errors on the data,
the RPI analysis gives an error of about 0.2\% for this ratio.
It is quite satisfactory that these \begin{it}two rather different
ways of analyzing the data agree so well on the extracted values
of these amplitudes.\end{it}

Comparing the results of these two solutions with
the excluded differential cross sections, general agreement is
found with two main exceptions: two sets 
of $\pi^0 p$ differential cross sections from Bonn \cite{gz74}, 
containing a total of 587 datum points. Thus, the $p\pi^0$ dataset
in this energy region is dominated by the Bonn datasets.
Therefore,
as an additional test of model dependence, and to determine the
influence of the Bonn data on the $M1$ and $E2$ amplitudes,
we have performed fits which
include these Bonn data, but still exclude all other pre-1980 differential
cross sections. The results, labelled as `Fit 2' are also shown in
Tables 1 and 2. Again, the results from RPI and VPI approaches
are in excellent agreement.
In fact, \begin{it}this is a very stringent test of model dependence on the
extracted $E2$ amplitude,\end{it} since the $E2$ amplitude is so small
in this case. Although at the T-matrix pole
the real and imaginary parts of this ratio have shifted compared to the
fit without the Bonn data, the magnitude is quite similar in both
fits.

Although the $M1$ amplitude is almost identical in the two fits,
the $E2/M1$ ratio is
quite sensitive to the chosen data set. This is particularly surprising
since the Mainz and Bonn $\pi^0$ differential cross sections are
in agreement.
The influence of these Bonn data on the extracted value of the $E2/M1$ ratio
has also been confirmed by the BNL group \cite{bnl}. 
In that work, as a test, the BNL cross sections are removed and 
replaced by the Bonn cross sections. The value of the extracted
EMR drops from -3.0\% to $-1.3$\%. Similar results have also been
found by the Mainz group \cite{lothar}. In the RPI fits, the raw data
have been fitted, that is, any systematic differences between different
datasets has not been taken into account. On the other hand, in the
VPI fits, the data were allowed to `float' in an attempt to take into
account systematic differences. As the effect of the Bonn data on the
$E2/M1$ ratio is found in both approaches, it seems to be a shape effect
rather than a scale effect.

Although the Mainz \cite{beck} and Bonn \cite{gz74} cross sections
agree over their common
angular range, their implications for the $E2$
amplitude are quite different.
Let us try to understand this.
Near the resonance energy, the Bonn 
cross sections \cite{gz74} range from 10$^\circ$ to 160$^\circ$ while the
Mainz cross sections \cite{beck} go from 75$^\circ$ to 125$^\circ$. Thus,
it is apparently
the Bonn data at forward and backward angles that are responsible
for bringing the $E2/M1$ ratio down in magnitude. As a check on this, we have
truncated the Bonn data to the same angular range as the Mainz data
and have redone the fit using the RPI approach. The result for the
$E2/M1$ ratio is -1.4\%, which accounts for much of the discrepancy
between the fits with and without the Bonn data.
It is also worth noting that the fit to the $\pi^0$
photon asymmetry data is significantly worsened when the Bonn data are
included in the fit, while the fit to all other observables remains largely
unchanged. Therefore, viewed from our two approaches,
\begin{it}there is an inconsistency between
the $\pi^0$ photon asymmetry data and the Bonn cross section data at
forward and backward angles.\end{it}

It is natural to ask how these two distinct observables could be in
disagreement. This is difficult to pin down quantitatively, because the
other multipoles, with the possible exception of the $M_{1+}^3$, will
differ if different datasets are fit. Qualitatively, however, the problem
seems to arise from the interference term between the $M_{1+}$ and
$E_{1+}$ multipoles, which appears both in the differential cross section and
the polarized photon asymmetry \cite{aron}.
The Bonn $\pi^0$ differential cross section
favors a small interference term, while the polarized photon
asymmetry favors a significantly larger value. The role of $E2$ in
these two observables can be judged by comparing the results of our
two fits, which are shown in Figs.~1 and 2 at 340 MeV.
The solid line is
the result from Fit 1, while the dashed line is obtained from Fit 2.
For the photon asymmetry (Fig.~1), Fit 1, which has the larger $E2$
amplitude, is clearly
favored, while for the differential cross section (Fig.~2), the backward angle
data from Bonn favor Fit 2. We can further investigate the role
of the $E2$ amplitude in these two observables by scaling
the $E2$ amplitude in Fit 2 such that the $E2/M1$
ratio is -3.2\% with everything else
held fixed. As is shown by the dotted line,
the agreement with the photon asymmetry is greatly improved when this
scaling is done, while the fit to the cross section becomes poorer.
Finally, it should be emphasized that the d waves
are allowed to vary in the VPI approach,
and therefore the discrepancy cannot be accounted for
by {\it these} multipoles.

In conclusion, we have demonstrated
that the RPI and VPI approaches give very similar
results for both the $M1$ and $E2$ amplitudes
in the case where all pre-1980 differential
cross sections are removed. These very different methods of analysis also 
agree on the effect of adding two sets \cite{gz74} of $\pi^0 p$ 
differential cross sections measured at Bonn in the 1970's. The agreement
between these two different approaches suggests that the model dependence
in extracting the {\it physical} resonant amplitudes is much smaller
than the range of $E2/M1$ values obtained from fitting different
databases. As our two approaches do not exhaust all the theoretical
methods used to analyze these data, a benchmark dataset, available to all
groups, would be useful for a broader investigation of the model
dependence in the extraction of the $M1$ and $E2$ amplitudes.

Looked at from our approaches, there is an inconsistency
between the Bonn $\pi^0 p$ differential cross section data at forward
and backward angles and the Mainz $\Sigma$ data. As the recent BNL $\Sigma$
data \cite{bnl} are consistent with the Mainz $\Sigma$ data,
it is vitally important to verify the forward and backward Bonn 
cross sections for neutral pion photoproduction. Given this,
the wider angle measurements of
the differential cross section for $\pi^0 p$ photoproduction from Mainz
\cite{beckpc}, not yet available in the current database,
could be crucial to a more definitive resolution of the $E2/M1$ problem,
of great topical interest to the understanding of baryon structure \cite{ncm}.

Although our main goal has been to investigate the model dependence in
the extraction of the $M1$ and $E2$ resonant amplitudes, some comments
on the preferred values of these amplitudes is in order. Of the datasets
considered here, the extracted value of the $M1$ amplitude
is quite stable with a
value of about 290$\times$10$^{-3}$ GeV$^{-1/2}$, which is roughly 30\%
larger than most quark model predictions \cite{quark}. On the other hand,
the recent BNL differential cross sections \cite{bnl} are larger than the 
previous Bonn and Mainz cross section measurements, presenting yet
another problem in the database. As the BNL cross sections imply an
$M1$ amplitude of about 310 in the same units, a resolution of the
discrepancy between these cross sections is urgently needed. Presently,
a conservative estimate of the $M1$ amplitude is 300$\pm$20, where
the error is almost entirely systematic.
For the $E2$ amplitude, or the $E2/M1$ ratio, a reasonable estimate
can be found
by examining the fit to the $\Sigma$ data at a photon lab-energy of 340
MeV, close to the K-matrix pole. Although most analyses have very similar
global chi-squares per degree of freedom, the quality of fit to these
particular data vary widely.
Comparing the results of various
multipole solutions with these data and discarding those that give poor
fits to these data, chi-squared per datum point greater than two,
one finds an $E2/M1$ range of about -2.5\% to -3.2\%. Thus, if
we adopt the $\Sigma$ data at 340 MeV as a benchmark,
then we have $E2/M1$ = -(2.8$\pm$0.4)\%.

An accurate
and consistent {\it complete set} of measurements is needed to resolve the
issues of the $M1$ and $E2$ nucleon to Delta(1232) transition amplitudes
once for all.

We thank R.~Beck, D.~Drechsel, A.~Sandorfi, P.~Stoler and L.~Tiator for
many spirited discussions and correspondences.
The RPI group is
supported by a U.S.~Department of Energy Grant No.~DE-FG02-88ER40448.A009,
while
the VPI group is supported in part by a U.S.~Department of Energy Grant
No.~DE-FG02-97ER41038. I.I.S.~is also supported in part
by a U.S.~Department of Energy Grant No.~DE-FG02-95ER40901. M.S.P.'s
research is a part of his undergraduate thesis at RPI, done under the
guidance of N.C.M.

\newpage
\begin{table}
\caption{Results of the K-matrix residues obtained from
fitting selected databases on pion photoproduction
observables in the Delta(1232) region by the RPI and VPI analyses.
Selection of the databases and these analyses are discussed in the text.
The K-matrix residues are in the standard units of 10$^{-3}$
GeV$^{-1/2}$.}
\label{table1}
\begin{center}
\begin{tabular}{|l|l|r|r|r|}\hline
\multicolumn{2}{|c|}{ } &
 \multicolumn{3}{c|}{K-matrix pole}  \\
 \cline{3-5}
\multicolumn{2}{|c|}{ } & $M1$ & $E2$ & $E2/M1$ \\
\hline
 & RPI& 289 & -6.0 & -2.1\%  \\
Fit 1 & &  &  &   \\
 & VPI& 290  & -5.6 & -1.9\%  
 \\ \hline
 & RPI& 290 & -1.3 & -0.45\% \\
Fit 2& &  &  &   \\
& VPI& 291  & -1.1 & -0.38\%  \\
\hline
\end{tabular}
\end{center}
\end{table}

\begin{table}
\caption{Results of T-matrix residues, in the standard units of 10$^{-3}$
GeV$^{-1/2}$, obtained from the RPI and VPI analyses.
For comparison with the Mainz group \protect\cite{mainz},
we also give $re^{i\phi}$ at the T-matrix pole, where
$r$ is in units of 10$^{-3}/\mu$ and $\phi$ is in degrees.}
\label{table2}
\begin{center}
\begin{tabular}{|l|l|c|c|c|c|c|c|c|}\hline
\multicolumn{2}{|c|}{ } &
 \multicolumn{7}{c|}{T-matrix pole} \\
 \cline{3-9}
\multicolumn{2}{|c|}{ }  & $M1$ & $E2$ & $E2/M1$ &
$r_M$ & $\phi_M$ & $r_E$ & $\phi_E$ \\
\hline
& RPI&  300+$i$27 &-8.7-$i$13.2 & -(3.3+$i$4.1)\% &
22.3 & -26.7 & 1.17 & -155.2 \\
Fit 1  &  &  & & & & & & \\
 & VPI&  297+$i$19 & -6.5-$i$15.9
 & -(2.5+$i$5.2)\% & 22.0 & -28.1 & 1.27 & -144.1
 \\ \hline
 & RPI&  301+$i$26 &-4.8-$i$14.8 & -(2.0+$i$4.7)\% &
22.3 & -26.9 & 1.15 & -139.8 \\
Fit 2  &  &  & & & & & & \\
& VPI&  301+$i$14  &-1.1-$i$15.0 & -(0.6+$i$4.9)\% &
22.3 & -29.1 & 1.11 & -126.0 \\
\hline
\end{tabular}
\end{center}
\end{table}

\noindent
Fig.~1: A comparison of the RPI results of Fit 1 (solid line) and
Fit 2 (dashed line), discussed in the text, with the recent Mainz
$\Sigma$ data \protect\cite{beck} at a photon lab-energy of 340 MeV.
The dotted line is obtained from Fit 2 by rescaling the $E2$ amplitude
such that the $E2/M1$ ratio is -3.2\%, everything else held fixed.

\vspace{24pt}
\noindent
Fig.~2: A comparison of the RPI results with the differential cross sections
from Mainz (diamonds) \protect\cite{beck} and Bonn (squares)
\protect\cite{gz74}. Curves as in Fig.~1.
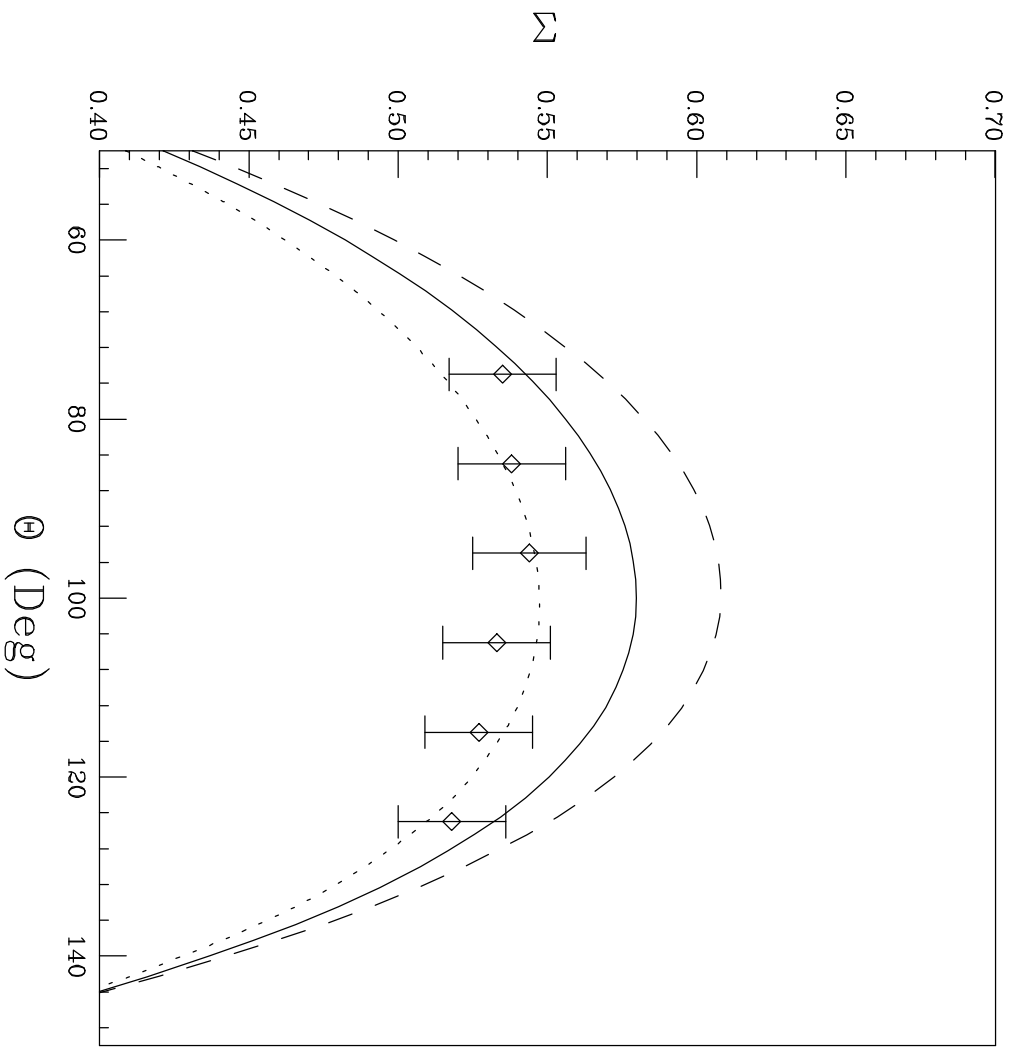
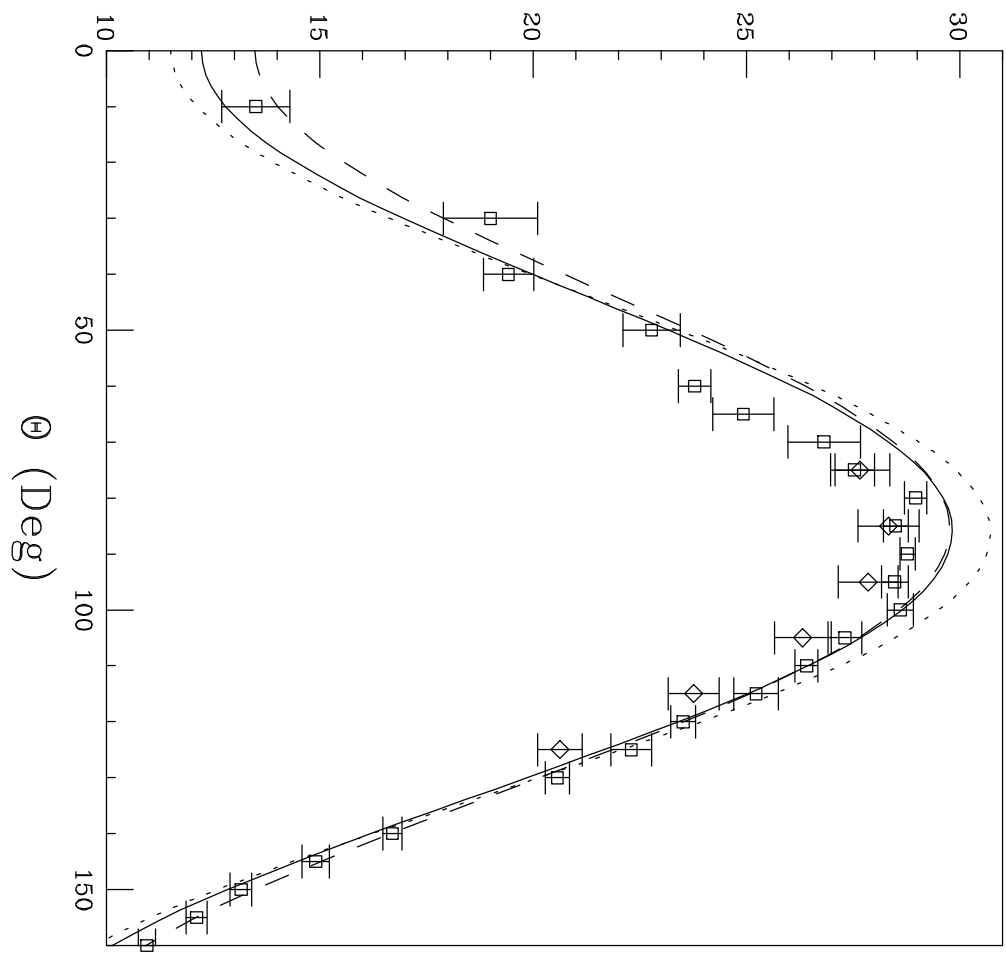
\end{document}